\documentclass{kapproc}
\usepackage[dvips]{graphicx}
\usepackage{procps}
\normallatexbib

\begin{document}
\articletitle[Entanglement of distant SQUID rings]{Photon-induced entanglement
of distant mesoscopic SQUID rings}
\author{A. Vourdas}
\affil{Department of Computing, University of Bradford, West Yorkshire, BD7
1DP, England.}
\author{D. I. Tsomokos}
\affil{Department of Computing, University of Bradford, West Yorkshire, BD7
1DP, England.}
\author{C. C. Chong}
\affil{Institute of High Performance Computing, 1 Science Park Road, 117 528,
Singapore.}

\abstract{An experiment that involves two distant mesoscopic SQUID rings is
studied. The superconducting rings are irradiated with correlated photons,
which are produced by a single microwave source. Classically correlated
(separable) and quantum mechanically correlated (entangled) microwaves are
considered, and their effect on the Josephson currents is quantified. It is
shown that the currents tunnelling through the Josephson junctions in the
distant rings, are correlated.}


\section{Introduction}
A fundamental property of superconducting quantum interference devices
(SQUIDs) is that they exhibit quantum coherence at the macroscopic level
\cite{1}. This property may be used for the purposes of quantum information
processing \cite{2,3}.

A lot of research on superconducting devices investigates their interaction
with classical microwaves. On the other hand the use of nonclassical
microwaves makes the system fully quantum mechanical and interesting quantum
phenomena arise. For example, in this paper we show that entangled two-mode
microwaves produce correlated currents in distant SQUID rings.

Nonclassical electromagnetic fields at low temperatures ($k_{\rm B}T \ll
\hbar\omega$) have been studied for more than twenty years both theoretically
and experimentally \cite{4}. The interaction of SQUID rings with nonclassical
microwaves has been studied in the literature \cite{5,6}.

In previous publications \cite{7} we have studied the effects of entangled
electromagnetic fields on distant electron interference experiments. In this
paper we review and extend further this work in the context of SQUID rings. We
consider two mesoscopic SQUID rings, which are far from each other and are
irradiated with entangled microwaves, produced by a single source (Fig. 1). It
is shown that the Josephson currents in the distant SQUID rings are
correlated. The photon-induced correlations between the currents are
quantified. It is shown that the current correlations depend on whether the
photons are classically correlated (separable) or quantum mechanically
correlated (entangled). The difference between separable and entangled
microwave density matrices \cite{8} is in the nondiagonal elements; and the
effect of these nondiagonal elements on the Josephson currents is explicitly
calculated.

\section{Interaction of a single SQUID ring with nonclassical microwaves}
In this section we consider a single SQUID ring and study its interaction with
both classical and nonclassical microwaves.

For irradiation with classical microwaves, the Josephson current is $I_{\rm
A}=I_1 \sin\theta_{\rm A}$, where $\theta_{\rm A}=2e\Phi_{\rm A}$ is the phase
difference across the junction due to the total flux $\Phi_{\rm A}$ through
the ring. We assume the external field approximation, where the back reaction
(the additional flux induced by the SQUID ring current) is neglected; i.e.,
the flux ${\cal L}I_{\rm A}$, where ${\cal L}$ is the self-inductance of the
ring, is negligible in comparison to $\Phi_{\rm A}$. The magnetic flux has a
linear and a sinusoidal component:
\begin{equation}
\Phi_{\rm A}=V_{\rm A}t+\phi_{\rm A};\;\;\;\;\;\phi_{\rm A}=A\sin(\omega_1 t).
\end{equation}
Consequently the observed current is
\begin{equation}
I_{\rm A}=I_1\sin[\omega_{\rm A}t+2eA\sin(\omega_1 t)];\;\;\;\;\;\omega_{\rm A}=2eV_{\rm A}.
\end{equation}
We now consider the interaction of a SQUID ring with nonclassical microwaves,
that are carefully prepared in a particular quantum state and are described by
a density matrix $\rho$. The dual quantum variables of the nonclassical field
are the vector potential $A_i$ and the electric field $E_i$. Integrating these
over the SQUID ring we obtain the magnetic flux and the electromotive force
operators $\hat\phi=\oint_C A_i dx_i$, $\hat V_{\rm EMF}=\oint_C E_i dx_i$.

In the external field approximation the flux operator evolves as
\begin{equation}
\hat\phi(t)=\xi 2^{-1/2}[\hat a^{\dagger}\exp(i\omega t)+
\hat{a}\exp(-i\omega t)],
\end{equation}
where $\xi$ is a parameter proportional to the area of the SQUID ring and the
$\hat{a}^{\dagger},\hat{a}$ are the photon creation and annihilation
operators. Consequently the phase difference $\theta_{\rm A}$ is the operator
\begin{eqnarray}\label{phase_theta_operator}
\hat{\theta}_{\rm A}=\omega_{\rm A}t + q[{\hat a}^{\dagger}\exp(i\omega t)+
\hat{a}\exp(-i\omega t)], \;\;\;\;\; q=\sqrt{2}e\xi;
\end{eqnarray}
and the current also becomes an operator, $\hat I_{\rm A}=I_1 \sin
\hat{\theta}_{\rm A}$. Expectation values of the current are calculated by
taking its trace with respect to the density matrix $\rho$, which describes
the nonclassical electromagnetic fields,
\begin{eqnarray}\label{I_A_expression}
\langle\hat I_{\rm A}\rangle &=& \mbox{Tr}(\rho \hat I_{\rm A})=
I_1 \mbox{Im} [\exp(i\omega_{\rm A}t)\tilde W(\lambda_{\rm A})], \\
\lambda_{\rm A}&=& iq\exp(i\omega_1 t).
\end{eqnarray}
$\tilde W(x)$ is the Weyl function \cite{9} given by
\begin{eqnarray}
\tilde W(x)=\mbox{Tr}[\rho D(x)];\;\;\;\;\;
D(x)=\exp(x \hat{a}^{\dagger}-x^{*}\hat{a})
\end{eqnarray}
where $D(x)$ is the displacement operator. Higher moments of the Josephson
current quantify the quantum statistics of the electron pairs tunnelling
through the junction.

\begin{figure}[h]
\begin{center}
\scalebox{0.3}{\includegraphics{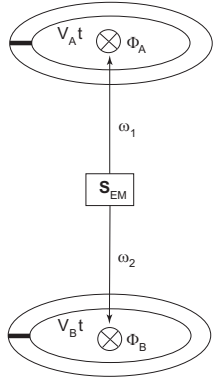}}
\end{center}
\caption{Two distant mesoscopic SQUID rings A and B are irradiated with
nonclassical microwaves of frequencies $\omega_1$ and $\omega_2$,
correspondingly. The microwaves are produced by the source ${\rm S}_{\rm EM}$
and are correlated. Classical magnetic fluxes $V_{\rm A}t$ and $V_{\rm B}t$
are also threading the two rings A and B, correspondingly.}
\end{figure}

\section{Interaction of two distant SQUID rings with entangled microwaves}
In this section we consider two mesoscopic SQUID rings far apart from each
other, which we refer to as A and B (Fig. 1). They are irradiated with
correlated microwaves. Let $\rho$ be the density matrix of the microwaves, and
$\rho_{\rm A}={\rm Tr}_{\rm B}\rho$, $\rho_{\rm B}={\rm Tr}_{\rm A}\rho$, the
density matrices of the microwaves interacting with the two SQUID rings ${\rm
A}$, ${\rm B}$, correspondingly. When the density matrix $\rho$ is
factorizable as $\rho_{\rm fact}=\rho_{\rm A}\otimes\rho_{\rm B}$ the two
modes are not correlated. If it can be written as  $\rho_{\rm sep}=\sum_i p_i
\rho_{{\rm A}i}\otimes \rho_{{\rm B}i}$, where $p_i$ are probabilities, it is
called separable and the two modes are classically correlated. Density
matrices which cannot be written in one of these two forms are entangled
(quantum mechanically correlated) \cite{8}.

The currents in the two SQUIDs are
\begin{eqnarray}\label{I_A}
\langle \hat I_{\rm A}\rangle = I_{1} \mbox{Tr}(\rho_{\rm A}\sin \hat
\theta_{\rm A}),  \;\;\;\;\; \langle \hat I_{\rm B}\rangle = I_{2}
\mbox{Tr}(\rho_{\rm B}\sin \hat \theta_{\rm B}).
\end{eqnarray}

The expectation value of the product of the two current operators is given by:
\begin{equation}\label{I_AB}
\langle\hat I_{\rm A}\hat I_{\rm B}\rangle = I_{1}I_{2} \mbox{Tr}(\rho
\sin\hat\theta_{\rm A}\sin\hat\theta_{\rm B}).
\end{equation}
In general $\langle\hat I_{\rm A}\hat I_{\rm B}\rangle$ is different from
$\langle\hat I_{\rm A}\rangle\langle\hat I_{\rm B}\rangle$
and we calculate the ratio
\begin{equation}\label{ratio}
R = \frac{\langle\hat I_{\rm A} \hat I_{\rm B}\rangle} {\langle\hat I_{\rm A}\rangle
\langle\hat I_{\rm B}\rangle}.
\end{equation}
For factorizable density matrices $\rho_{\rm fact}=\rho_{A}\otimes\rho_B$ we
easily see that $R_{\rm fact}=1$. For separable density matrices $\rho_{\rm
sep}$ the ratio $R_{\rm sep}$ is not necessarily equal to one and numerical
results for various examples are shown below.

We also calculate the second moments
\begin{eqnarray}
\langle {\hat I}_{\rm A}^2 \rangle = I_{1}^2 \mbox{Tr}[\rho_{\rm A}(\sin
\hat\theta_{\rm A})^2], \;\;\;\;\; \langle {\hat I}_{\rm B}^2\rangle = I_{2}^2
\mbox{Tr}[\rho_{\rm B}(\sin \hat \theta_{\rm B})^2]. \label{I_A_2}
\end{eqnarray}
The statistics of the photons affects the statistics of the tunnelling
electron pairs, which is quantified with the $\langle {\hat I}_{\rm A} {\hat
I}_{\rm B}\rangle$, $\langle {\hat I} _{\rm A}^{2} \rangle $, $\langle {\hat
I}^{2}_{\rm B}\rangle$ (and also with the higher moments).

\subsection{Microwaves in number states}
We consider microwaves in the separable (mixed) state
\begin{equation}\label{num_1_sep}
\rho_{\rm sep}=\frac{1}{2}(|N_1 N_2 \rangle \langle N_1 N_2| + |N_2N_1\rangle
\langle N_2N_1|),
\end{equation}
where $N_1\neq N_2$. We also consider microwaves in the entangled state
$|s\rangle=2^{-1/2}(|N_1N_2\rangle+|N_2N_1\rangle)$, which is a pure state.
The density matrix of $|s\rangle$ is
\begin{equation}\label{num_1_ent}
\rho_{\rm ent}=\rho_{\rm sep} + \frac{1}{2}(|N_1 N_2 \rangle \langle N_2
N_1|+|N_2 N_1 \rangle \langle N_1 N_2|),
\end{equation}
where the $\rho_{\rm sep}$ is given by Eq. (\ref{num_1_sep}). It is seen that
the $\rho_{\rm ent}$ and the $\rho_{\rm sep}$ differ only by the above
nondiagonal elements.

In this example, the reduced density matrices are the same for both the
separable and entangled states:
\begin{eqnarray}\label{reduced_rho}
\rho_{\rm sep, A}=\rho_{\rm ent, A}=\rho_{\rm sep, B}=\rho_{\rm ent, B} =
\frac{1}{2} (|N_1\rangle\langle N_1| + |N_2\rangle\langle N_2|).
\end{eqnarray}
Consequently in this example $\langle\hat I_{\rm A}\rangle_{\rm
sep}=\langle\hat I_{\rm A}\rangle_{\rm ent}$, and also $\langle\hat I_{\rm
B}\rangle_{\rm sep}=\langle\hat I_{\rm B}\rangle_{\rm ent}$.

For the density matrix $\rho_{\rm sep}$ of Eq. (\ref{num_1_sep}) we find
\begin{eqnarray}
\langle\hat I_{\rm A}\rangle &=& \frac{I_{1}}{2}
\exp\left(-\frac{q^2}{2}\right)[L_{N_1}(q^2)+L_{N_2}(q^2)]
\sin(\omega_{\rm A}t), \label{I_reduced_A_num}\\
\langle\hat I_{\rm B}\rangle &=&
\frac{I_{2}}{2}\exp\left(-\frac{q^2}{2}\right)[L_{N_1}(q^2)+L_{N_2}(q^2)]
\sin(\omega_{\rm B}t), \label{I_reduced_B_num}
\end{eqnarray}
where the $L_{n}^{\alpha}(x)$ are Laguerre polynomials. The currents
$\langle\hat I_{\rm A}\rangle, \langle\hat I_{\rm B}\rangle$ are in this
example independent of the microwave frequencies $\omega_1,\omega_2$.

The expectation value of the product of the two currents [Eq. (\ref{I_AB})] is
\begin{eqnarray}
\langle{\hat I}_{\rm A} {\hat I}_{\rm B}\rangle_{\rm sep}&=& I_{1}I_{2}
\exp(-q^2) L_{N_1}(q^2)L_{N_2}(q^2) \sin(\omega_{\rm A}t)\sin(\omega_{\rm
B}t). \label{I_sep}
\end{eqnarray}
Consequently the ratio $R$ of Eq. (\ref{ratio}) is
\begin{equation}\label{R_sep}
R_{\rm sep}= \frac{4L_{N_1}(q^2)L_{N_2}(q^2)} {[L_{N_1}(q^2)+L_{N_2}(q^2)]^2}.
\end{equation}
In this example the $R_{\rm sep}$ is time-independent.

The moments of the currents, defined by Eq. (\ref{I_A_2}), are also
calculated:
\begin{eqnarray}
\langle \hat I_{\rm A}^2 \rangle &=&  \frac{I_{1}^2}{2} \left\{1 -
\frac{1}{2}\exp(-2q^2)[L_{N_1}(4q^2)+L_{N_2}(4q^2)]
 \cos(2\omega_{\rm A}t) \right\}, \label{I_A_squared} \\
\langle \hat I_{\rm B}^2 \rangle &=&  \frac{I_{2}^2}{2} \left\{1 -
\frac{1}{2}\exp(-2q^2)[L_{N_1}(4q^2)+L_{N_2}(4q^2)] \cos(2\omega_{\rm B}t)
\right\}. \label{I_B_squared}
\end{eqnarray}

For the case of $\rho_{\rm ent}$ the $\langle\hat I_{\rm A}\rangle,\langle\hat
I_{\rm B}\rangle$ are the same as in Eqs. (\ref{I_reduced_A_num}),
(\ref{I_reduced_B_num}); and the $\langle\hat I_{\rm A}^2\rangle,\langle\hat
I_{\rm B}^2\rangle$ are the same as in Eqs. (\ref{I_A_squared}),
(\ref{I_B_squared}). However the $\langle{\hat I}_{\rm A}{\hat I}_{\rm
B}\rangle$ is
\begin{equation}\label{I_ent}
\langle{\hat I}_{\rm A}{\hat I}_{\rm B}\rangle_{\rm ent} = \langle{\hat
I}_{\rm A}{\hat I}_{\rm B}\rangle_{\rm sep} + I_{\rm cross},
\end{equation}
where
\begin{eqnarray}
I_{\rm cross} &=& - \frac{I_{1}I_{2}}{2} \exp(-q^2)
L_{N_1}^{N_2-N_1}(q^2)L_{N_2}^{N_1-N_2}(q^2)
[\cos(\omega_{\rm A}t+\omega_{\rm B}t) \nonumber \\
&& -(-1)^{N_1-N_2} \cos(\omega_{\rm A}t-\omega_{\rm B}t)] \cos(\Omega t), \label{I_cross} \\
\Omega &=& (N_1-N_2)(\omega_1-\omega_2).
\end{eqnarray}
It is seen that the effect of entangled microwaves on Josephson currents is
different from the effect of separable  microwaves. In this case the ratio $R$
of Eq. (\ref{ratio}) is
\begin{equation}
R_{\rm ent} = R_{\rm sep} + \frac{I_{\rm cross}(t)}{\langle \hat{I}_{\rm
A}\rangle \langle \hat{I}_{\rm B}\rangle},
\end{equation}
which is a time-dependent quantity oscillating around the $R_{\rm sep}$.

\subsection{Microwaves in coherent states}
We consider microwaves in the classically correlated state
\begin{eqnarray}\label{rho_sep_coherent}
\rho_{{\rm sep}}=\frac{1}{2}(|A_1 A_2\rangle \langle A_1 A_2| +|A_2 A_1\rangle
\langle A_2 A_1|),
\end{eqnarray}
where $|A_1\rangle$, $|A_2\rangle$ are microwave coherent states. We also
consider the entangled state $|u\rangle={\cal N}(|A_1 A_2\rangle +|A_2
A_1\rangle)$, with density matrix
\begin{eqnarray}\label{rho_ent_coherent}
\rho_{\rm ent}= 2{\cal N} ^2 \rho_{\rm sep}+ {\cal N}^2 (|A_1 A_2\rangle
\langle A_2 A_1| +|A_2 A_1\rangle \langle A_1 A_2|),
\end{eqnarray}
where the normalization constant is given by
\begin{eqnarray}\label{normalization_N}
{\cal N}=\left[2+2\exp\left(-|A_1-A_2|^2\right)\right]^{-1/2}.
\end{eqnarray}

For microwaves in the separable state of Eq. (\ref{rho_sep_coherent}) the
reduced density matrices are
\begin{eqnarray}\label{reduced_rho_sep}
\rho_{\rm sep,A}=\rho_{\rm sep,B}=\frac{1}{2}(|A_1\rangle \langle A_1|
+|A_2\rangle\langle A_2|),
\end{eqnarray}
and hence the current in A is
\begin{eqnarray} \label{I_A_sep}
\langle\hat I_{\rm A}\rangle_{\rm sep} &=& \frac{I_{1}}{2}\exp(-\frac{q^2}{2})
\{\sin[\omega_{\rm A}t+2q|A_1|\cos(\omega_1 t-\theta_{1})] \nonumber \\
&+& \sin[\omega_{\rm A}t+2q|A_2|\cos(\omega_1 t-\theta_{2})]\}.
\end{eqnarray}
where $\theta_1=\arg (A_1)$, and $\theta_2=\arg (A_2)$. A similar expression
yields the current in B. We have also calculated numerically the ratio $R_{\rm
sep}$.

\begin{figure}[h]
\begin{center}
\scalebox{0.4}{\includegraphics{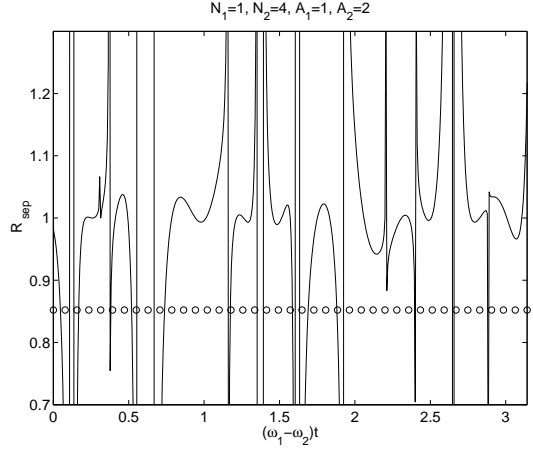}}
\end{center}
\caption{$R_{\rm sep}$ against $(\omega_1-\omega_2)t$ for the number state of
Eq. (\ref{num_1_sep}) with $N_1=1,N_2=4$ (line of circles), and the coherent
state of Eq. (\ref{rho_sep_coherent}) with $A_1=1,A_2=2$ (solid line). The
photon frequencies are $\omega_1=1.2\times 10^{-4}$ and $\omega_2=10^{-4}$, in
units where $k_B=\hbar=c=1$.}
\end{figure}

For microwaves in the entangled state of Eq. (\ref{rho_ent_coherent}) the
reduced density matrices are
\begin{eqnarray}\label{reduced_rho_ent}
\rho_{\rm ent,A}=\rho_{\rm ent,B}= {\cal N}^2(|A_1\rangle \langle A_1|
+|A_2\rangle \langle A_2| + \tau |A_1\rangle\langle A_2| + \tau^{*}
|A_2\rangle\langle A_1|),
\end{eqnarray}
where $\tau = \langle A_1|A_2\rangle = \exp(-|A_1|^2/2 -|A_2|^2/2 + A_1^{*}
A_2)$.

The current in A is
\begin{eqnarray}
\langle\hat I_{\rm A}\rangle_{\rm ent}= 2{\cal N}^2 \langle\hat I_{\rm
A}\rangle_{\rm sep}+ {\cal N}^2 E F_1 \exp\left(-\frac{q^2}{2}\right)I_{1},
\end{eqnarray}
where $E = \exp[-|A_1|^2 - |A_2|^2+2|A_1 A_2|\cos(\theta_{1}-\theta_{2})]$,
and
\begin{eqnarray}
\lefteqn{ F_1 = [\exp(q|A_1|S_{A,1}-q|A_2|S_{A,2})+ \exp(-q|A_1|S_{A,1}} \nonumber \\
&& +q|A_2|S_{A,2})] \sin(\omega_A t+q|A_1|
C_{A,1}+q|A_2|C_{A,2}),\label{I_A_ent}
\end{eqnarray}
with $S_{A,1}=\sin(\omega_1 t-\theta_{1})$, $S_{A,2}=\sin(\omega_1
t-\theta_{2})$, $C_{A,1}=\cos(\omega_1 t-\theta_{1})$,
$C_{A,2}=\cos(\omega_1t-\theta_{2})$. A similar expression yields the current
in B, and we have also calculated numerically the ratio $R_{\rm ent}$.

\begin{figure}[h]
\begin{center}
\scalebox{0.4}{\includegraphics{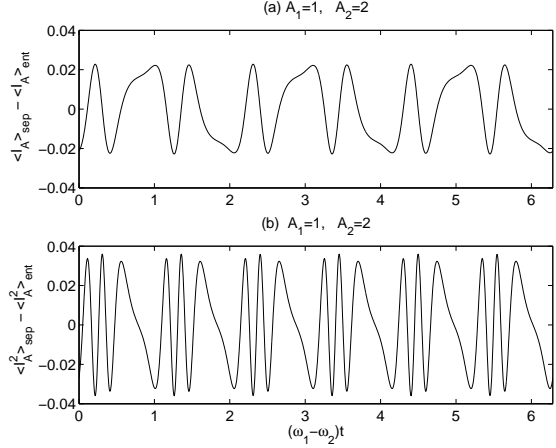}}
\end{center}
\caption{(a) $\langle {\hat I}_{\rm A}\rangle_{\rm sep} - \langle {\hat
I}_{\rm A} \rangle_{\rm ent}$, and (b) $\langle {\hat I}_{\rm A}^2
\rangle_{\rm sep} - \langle {\hat I}_{\rm A}^2 \rangle_{\rm ent}$ against
$(\omega_1-\omega_2)t$ for the coherent state $\rho_{\rm sep,A}$ of Eq.
(\ref{reduced_rho_sep}) and $\rho_{\rm ent,A}$ of Eq. (\ref{reduced_rho_ent})
with $A_1=1,A_2=2$. The photon frequencies are $\omega_1=1.2\times 10^{-4}$
and $\omega_2=10^{-4}$, in units where $k_B=\hbar=c=1$.}
\end{figure}

\subsection{Numerical results}
In the numerical results of Figs. 2 and 3 the microwave frequencies are
$\omega_1=1.2\times 10^{-4},\omega_2=10^{-4}$, in units where $k_B=\hbar=c=1$.
The critical currents are $I_{1}=I_{2}=1$. The other parameters are $\xi=1$,
$\omega_{\rm A}=\omega_1$, $\omega_{\rm B}=\omega_2$, $N_1=1$, $N_2=4$, and
$\theta_1=\theta_2=0$. For a meaningful comparison between microwaves in
number states and microwaves in coherent states, we take them to have the same
average number of photons, $|A_1|^2=N_1$ and $|A_2|^2=N_2$.

In Fig. 2 we plot $R_{\rm sep}$ against $(\omega_1-\omega_2)t$ for currents
induced by microwaves in the number state of Eq. (\ref{num_1_sep}) with
$N_1=1,N_2=4$ (line of circles), and the coherent state of Eq.
(\ref{rho_sep_coherent}) with $A_1=1,A_2=2$ (solid line). It is seen that two
different separable photon states with the same average number of photons give
rise to different correlations between the SQUID currents.

In Fig. 3 we plot (a) $\langle {\hat I}_{\rm A}\rangle_{\rm sep} - \langle
{\hat I}_{\rm A} \rangle_{\rm ent}$, and (b) $\langle {\hat I}_{\rm A}^2
\rangle_{\rm sep} - \langle {\hat I}_{\rm A}^2 \rangle_{\rm ent}$, against
$(\omega_1-\omega_2)t$ for microwaves in the coherent state $\rho_{\rm A,sep}$
of Eq. (\ref{reduced_rho_sep}) and $\rho_{\rm A,ent}$ of Eq.
(\ref{reduced_rho_ent}) with $A_1=1,A_2=2$. In Fig. 3(a) it is seen that the
Josephson current in SQUID ring A is different for irradiation with separable
and entangled microwaves in coherent states. In Fig. 3(b) it is seen that
irradiation with separable and entangled coherent states leads to different
second moments of the current, which implies that the quantum statistics of
electron pairs tunnelling the Josephson junction of SQUID ring A are different
in these two cases.

\section{Discussion}
We have considered the interaction of two distant SQUID rings A and B with
two-mode nonclassical microwaves, which are produced by the same source. The
flux, the phase difference and the Josephson currents are operators and their
expectation values with the density matrix of the nonclassical microwaves give
the physically observed quantities. We have assumed the external field
approximation, where the electromagnetic field created by the Josephson
currents (back reaction) is neglected and we have calculated various
quantities.

It has been shown that the Josephson currents in the two rings are correlated
in the sense that $\langle\hat I_{\rm A}\hat I_{\rm B}\rangle$ is different
from $\langle\hat I_{\rm A}\rangle\langle\hat I_{\rm B}\rangle$ (for
non-factorizable density matrices). We have considered examples where the
photons are classically correlated and quantum mechanically correlated; and we
have shown that the non-diagonal terms in the latter case affect the Josephson
currents. Further work in this direction could be the formulation of Bell-type
inequalities for the Josephson currents, which are obeyed in the case of
separable microwaves and violated in the case of entangled microwaves.


\end{document}